\documentclass[floats,aip,cha,reprint,superscriptaddress]{revtex4-1}

\usepackage{graphicx}
\usepackage{amsmath}   
\usepackage{amsfonts}
\usepackage{nicefrac}   
\usepackage{enumerate}

\providecommand*{\p}{\ensuremath{p}}
\providecommand*{\q}{\ensuremath{q}}

\providecommand*{\xic}{\ensuremath{\xi_{12}}}

\providecommand*{\pointPSname}{\ensuremath{u}}
\providecommand*{\pointPS}{\ensuremath{\vec{\pointPSname}}}

\providecommand*{\fp}{\ensuremath{\mathrm{fp}}}

\providecommand*{\periodicpoint}{\ensuremath{{\pointPS_{\mathrm{p}}}}}
\providecommand*{\alphapoint}{\ensuremath{{\pointPS_{\alpm{2}}}}}
\providecommand*{\fixedpoint}{\ensuremath{{\pointPS_{\fp}}}}
\providecommand*{\valfixpt}[1]{\ensuremath{{\nu_{#1}^\fp}}}
\providecommand*{\vecfixpt}[2]{\ensuremath{{\vec{v}_{#1#2}^\fp}}}

\providecommand*{\oneD}{\textsc{1d}}
\providecommand*{\twoD}{\textsc{2d}}
\providecommand*{\threeD}{\textsc{3d}}
\providecommand*{\fourD}{\textsc{4d}}

\providecommand*{\mayavi}{\textsc{Mayavi}}

\providecommand*{\nua}{\ensuremath{\valfixpt{1}}}
\providecommand*{\nub}{\ensuremath{\valfixpt{2}}}

\providecommand*{\C}{\ensuremath{{\cal C}}}
\providecommand*{\Ca}{\ensuremath{{\cal C}_1^{\fp}}}
\providecommand*{\Cb}{\ensuremath{{\cal C}_2^{\fp}}}
\providecommand*{\Cpo}[1]{\ensuremath{{\cal C}_{#1}^{7}}}
\providecommand*{\Cres}{\ensuremath{{\cal C}^{\mathrm{res}}}}
\providecommand*{\Caa}[1]{\ensuremath{{\cal C}_{#1}^{\alpm{2}}}}

\providecommand*{\M}{\ensuremath{{\cal M}}}
\providecommand*{\Ma}{\ensuremath{{\cal M}_1^{\fp}}}
\providecommand*{\Mb}{\ensuremath{{\cal M}_2^{\fp}}}
\providecommand*{\Mpo}[1]{\ensuremath{{\cal M}_{#1}^{7}}}
\providecommand*{\Mres}{\ensuremath{{\cal M}^{\mathrm{res}}}}
\providecommand*{\Maa}[1]{\Mh{\alpm{2}}{#1}}
\providecommand*{\Mh}[2]{\ensuremath{{\cal M}^{#1}_{#2}}}
\providecommand*{\Th}[2]{\ensuremath{{\cal T}^{#1}_{#2}}}

\providecommand*{\sectioneps}{\ensuremath{\varepsilon}}

\providecommand*{\symbolsection}{\ensuremath{\Gamma}}
\providecommand*{\Geps}[1][\sectioneps]{\ensuremath{\symbolsection_{#1}}}

\providecommand*{\coloralphaa}{turquoise}
\providecommand*{\coloralphab}{green}

\providecommand*{\colortwotorus}{gray}
\providecommand*{\colorcentera}{orange}
\providecommand*{\colorcenterb}{red}

\providecommand*{\colorperiodseven}{blue}

\providecommand*{\colorresonancethreetower}{magenta}

\providecommand*{\colorthreetower}{\colorresonancethreetower}

\providecommand*{\psslc}{\threeD{} phase-space slice}
\providecommand*{\psslcs}{\psslc s}

\providecommand*{\alpm}[1]{\ensuremath{\alpha #1}}
\providecommand*{\betm}[1]{\ensuremath{\beta #1}}
\providecommand*{\alp}[1]{(\alpm{#1})}
\providecommand*{\bet}[1]{(\betm{#1})}

\providecommand*{\subfiga}{(a)}
\providecommand*{\subfigb}{(b)}
\providecommand*{\subfigc}{(c)}
\providecommand*{\subfigd}{(d)}
\providecommand*{\subfige}{(e)}
\providecommand*{\subfigf}{(f)}

\ifx\theResA\undefined
\newcounter{ResA}
\fi
\ifx\theResE\undefined
\newcounter{ResE}
\fi

\newlength{\subfigurewidth}
\usepackage{array}
\newcolumntype{P}[1]{>{\centering\let\newline\\\arraybackslash\hspace{0pt}}p{#1}}
\newcolumntype{M}[1]{>{\centering\let\newline\\\arraybackslash\hspace{0pt}}m{#1}}

\newcolumntype{L}[1]{>{\raggedleft\let\newline\\\arraybackslash\hspace{0pt}}b{#1}}

\usepackage{amsfonts}

\providecommand*{\N}{\mathbb{N}}

\providecommand*{\C}{\mathbb{C}}

\newcounter{resstructcolumncounter}

\providecommand*{\ue}{\text{e}}
\providecommand*{\ui}{\text{i}}
\usepackage{color}

\newcommand{\MOVIEREF}{%
  For a rotating view see~\cite{suppMat}}

\usepackage{hyperref}

\begin{document}

\title{Global structure of regular tori in a generic 4D symplectic map}

\author{S.\ Lange}
\affiliation{Technische Universit\"{a}t Dresden, Institut f\"{u}r
  Theoretische Physik and Center for Dynamics, 01062 Dresden, Germany}
\affiliation{Max-Planck-Institut f\"{u}r Physik komplexer Systeme,
  N\"{o}thnitzer Stra{\ss}e 38, 01187 Dresden, Germany}

\author{M.\ Richter}
\affiliation{Technische Universit\"{a}t Dresden, Institut f\"{u}r
  Theoretische Physik and Center for Dynamics, 01062 Dresden, Germany}
\affiliation{Max-Planck-Institut f\"{u}r Physik komplexer Systeme,
  N\"{o}thnitzer Stra{\ss}e 38, 01187 Dresden, Germany}

\author{F. Onken}
\affiliation{Technische Universit\"{a}t Dresden, Institut f\"{u}r
  Theoretische Physik and Center for Dynamics, 01062 Dresden, Germany}

\author{A.\ B\"acker}
\affiliation{Technische Universit\"{a}t Dresden, Institut f\"{u}r
  Theoretische Physik and Center for Dynamics, 01062 Dresden, Germany}
\affiliation{Max-Planck-Institut f\"{u}r Physik komplexer Systeme,
  N\"{o}thnitzer Stra{\ss}e 38, 01187 Dresden, Germany}

\author{R.\ Ketzmerick}
\affiliation{Technische Universit\"{a}t Dresden, Institut f\"{u}r
  Theoretische Physik and Center for Dynamics, 01062 Dresden, Germany}
\affiliation{Max-Planck-Institut f\"{u}r Physik komplexer Systeme,
  N\"{o}thnitzer Stra{\ss}e 38, 01187 Dresden, Germany}

\date{\today}

\begin{abstract}
  For the case of generic \fourD{} symplectic maps with a mixed phase
  space we investigate the global organization of regular tori. For
  this we compute elliptic $1$-tori of two coupled standard maps and
  display them in a \psslc{}. This visualizes how all regular $2$-tori
  are organized around a skeleton of elliptic $1$-tori in the \fourD{}
  phase space. The $1$-tori occur in two types of one-parameter
  families: \alp{} Lyapunov families emanating from
  elliptic-elliptic periodic orbits, which are observed to
  exist even far away from them and beyond major
  resonance gaps, and \bet{} families originating from
  rank-$1$ resonances. At resonance gaps of both types of families
  either (i) periodic orbits exist, similar to the Poincar\'e-Birkhoff
  theorem for \twoD{} maps, or (ii) the family may form large bends.
  In combination these results allow for describing the
  hierarchical structure of regular tori in the \fourD{} phase space analogously to
  the islands-around-islands hierarchy in \twoD{} maps.
\end{abstract}


\maketitle

\noindent


\begin{quotation}
  For Hamiltonian systems with two degrees of freedom or
  two-dimensional area-preserving maps a detailed understanding of
  dynamics is well-established: Around stable periodic orbits one
  typically has invariant regular tori if their frequency is
  sufficiently irrational as predicted by the Kolmogorov-Arnold-Moser
  (KAM) theorem. For rational frequencies, the Poincar\'e-Birkhoff
  theorem predicts a chain of elliptic and hyperbolic periodic orbits.
  These structures can be directly visualized by two-dimensional plots
  of the dynamics providing a high level of intuition. In this paper
  we make progress towards a similar level of understanding for
  higher-dimensional systems using 3D phase-space slices to visualize
  the dynamics. For a 4D map we show that all regular 2-tori are
  organized around a skeleton of elliptic 1-tori, see
  Fig.~\ref{fig:psslc-central-tori}. These are associated either with
  elliptic-elliptic periodic orbits or with rank-1 resonances. The
  visualization allows for an intuitive understanding of the
  organization of regular tori and their hierarchy in
  higher-dimensional systems analogously to 2D maps.
\end{quotation}

\section{\label{sec:intro}Introduction}

\begin{figure}[b]
  \includegraphics{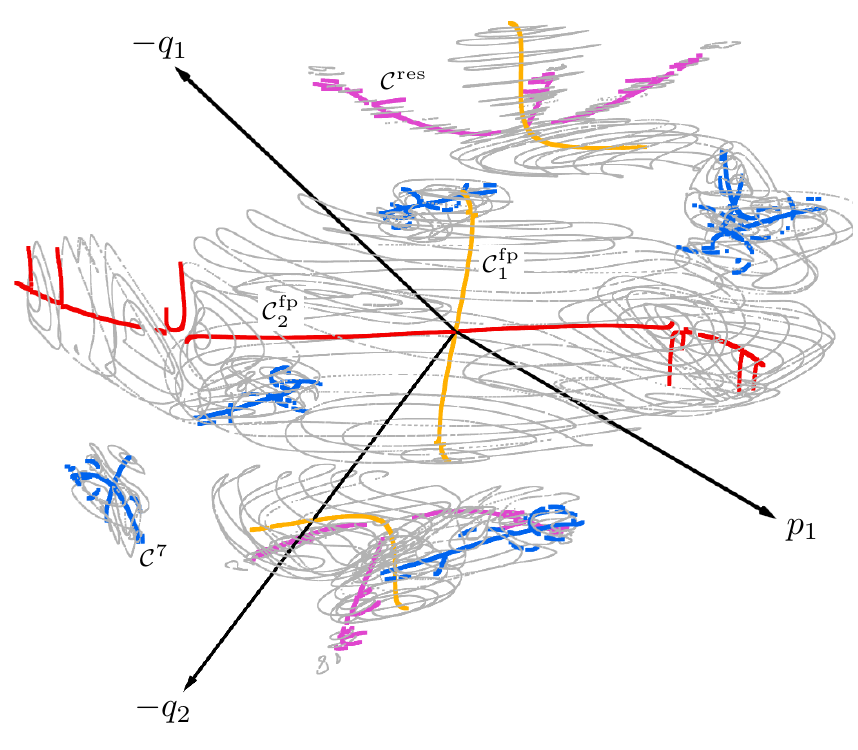}
  \caption{\label{fig:psslc-central-tori}%
    Central lines \Ca, \Cb, \Cpo, \Cres{} (\colorcentera,
    \colorcenterb, \colorperiodseven, \colorthreetower) in the
    \psslc{} representing families of $1$-tori of type \alp{} \Ma,
    \Mb, \Mpo, and type \bet{} \Mres{} for two coupled standard maps,
    Eq.~(\ref{eq:CoupledStdMaps}). Around the central lines the
    $2$-tori (\colortwotorus) are organized. From the
    elliptic-elliptic fixed point \fixedpoint{} the two central lines
    \Ca, \Cb{} (\colorcentera, \colorcenterb) emanate, and continue
    beyond the large gaps. The central lines \Cres{}
    (\colorthreetower{}) arise from a $-1:3:0$ resonance. \MOVIEREF}
\end{figure}

The dynamics of higher-dimensional Hamiltonian systems is intensively
studied in many areas of physics, chemistry, and
mathematics~\cite{LicLie1992}, such as the solar
system~\cite{UdrPfe1988,Las1990,Cin2002}, particle
accelerators~\cite{DumLas1993}, atoms and
molecules~\cite{RicWin1990,SchBuc1999,GekMaiBarUze2006,Kes2007,PasChaUze2008,WaaSchWig2008},
as well as KAM-theory~\cite{Gra1974,Zeh1976,Poe1989,JorVil1997b}.
Especially important phenomena are Arnold
diffusion~\cite{Arn1964,Chi1979,LicLie1992,Cin2002}, which exclusively
occurs in higher-dimensional systems, and power-law trapping or stickiness of
chaotic
orbits~\cite{DinBouOtt1990,AltMotKan2006,AltKan2007,She2010,Lan2012},
whose mechanism is still unknown for these systems. To investigate
these questions it is particularly convenient to study
$2f$-dimensional symplectic maps, which, e.g., arise from autonomous
Hamiltonian systems with $f+1$ degrees of freedom or time-periodically
driven systems with $f$ degrees of freedom.

For \twoD{} symplectic maps the organization of structures in phase
space is well understood~\cite{LicLie1992}: Around elliptic fixed points
and periodic orbits one has \oneD{} regular tori with sufficiently
irrational frequencies as predicted by the Kolmogorov-Arnold-Moser
(KAM) theorem. This surrounding of an elliptic point is referred to as
a regular island. In contrast, resonant regular tori break up in the
non-integrable case into chains of elliptic and hyperbolic periodic
orbits according to the Poincar\'e-Birkhoff theorem. This gives rise
to a whole hierarchy of regular islands, e.g.,
islands-around-islands~\cite{Mei1986}. This hierarchy along with its
affiliated structures, like stable and unstable manifolds, governs the
dynamics in phase space.

For higher-dimensional maps the organization of regular tori in phase
space is more complicated: For a $2f$-dimensional symplectic map the
corresponding phase space in general contains regular tori of
dimensions $d = 0,\ldots, f$~\cite{Gra1974,Zeh1976,Eli1988,Poe1989},
in the following denoted as $d$-tori. The behavior normal to
lower-dimensional tori ($0<d<f$) and fixed points ($d=0$) can be an
arbitrary combination of hyperbolic and elliptic
components~\cite{Sev1997,JorVil1997,JorOll2004}. A $d$-torus is called
elliptic or hyperbolic if all its normal components are elliptic or
hyperbolic~\cite{Sev1997}. Hyperbolic tori, also called whiskered
tori, are important for many dynamical properties of a system,
including Arnold
diffusion~\cite{Arn1964,Gra1974,Zeh1976,Sev2008,WaaSchWig2008}. The
regular tori have a hierarchical ordering in the sense that Cantor
families of elliptic $d$-tori are arranged around elliptic
$(d-1)$-tori for $0<d\leq f$ in an intricate
way~\cite{Sev1997,JorVil1997,JorVil2001}. In particular, the elliptic
(lower-dimensional) tori are investigated in the context of
KAM-theory~\cite{Tod1994,JorVil1997b,JorVil2001,Sev2008,LuqVil2011},
important for the description of
(Hopf)-bifurcations~\cite{Pfe1985a,RoyLah1991,LahBhoRoy1998,JorOll2004},
and one origin of stickiness~\cite{MorGio1995, AltMotKan2005,
  AltMotKan2006, Bun2008}. Furthermore, the hierarchical ordering of
elliptic tori has been studied for the hydrogen atom in crossed
fields~\cite{GekMaiBarUze2006,GekMaiBarUze2007} and in the context of
Hamilton-Hopf bifurcations~\cite{SevLah1991,JorOll2004}, and observed
in solar systems~\cite{VraIslBou1997, CouLasCorMayUdr2010}. However,
despite the available analytical and numerical results a complete
description of the phase-space structures and their geometry is still
missing for higher-dimensional systems. Most of the mentioned results
have been obtained by normal form
tools~\cite{Gra1974,Zeh1976,Eli1988,Poe1989,
  Tod1994,JorVil1997b,JorVil1997} or other perturbative
schemes~\cite{SevLah1991,RoyLah1991,LahBhoRoy1998} and therefore only
cover near-integrable systems or the vicinity of invariant objects,
e.g., fixed points. However, many practical applications are concerned
with non-perturbative
systems~\cite{GekMaiBarUze2006,GekMaiBarUze2007,PasChaUze2008,LuqVil2011}.
Thus, it is particularly relevant to investigate the phase-space
structures in systems far from integrability and far from fixed
points.

The aim of this paper is to investigate and visualize the organization
of regular tori of generic \fourD{} symplectic maps to gain an
understanding similar to the insights already available for \twoD{}
maps. For this we use a new numerical algorithm to compute elliptic
$1$-tori from their surrounding $2$-tori. We employ the
\psslcs{}~\cite{RicLanBaeKet2014} to simultaneously display both $1$-
and $2$-tori, see Fig.~\ref{fig:psslc-central-tori}. The \threeD{}
impression is considerably enhanced when one rotates the figure with
standard \threeD{} graphics. For the convenience of the reader for all
\psslcs{} in this paper videos with rotating camera position are given
in the supplemental material~\cite{suppMat}. The \psslcs{} clearly visualize that
families of elliptic $1$-tori form a skeleton for the surrounding
$2$-tori. We show that two types of such families can be distinguished
depending on whether they are associated with \alp{} an
elliptic-elliptic fixed point (or elliptic-elliptic periodic orbit) or
\bet{} a rank-$1$ resonance. The former families originate from the
center submanifolds of the elliptic-elliptic fixed points and are also
called \emph{Lyapunov families of invariant
  curves}~\cite{JorVil1997b,JorOll2004}, while the latter are
fundamentally different. We explain the origin of this type of
families in detail based on results concerning the break-up of
resonant $2$-tori~\cite{Tod1994} and frequency
analysis~\cite{Las1993}. Furthermore, we find two kinds of resonance
gaps which interrupt the skeleton of elliptic $1$-tori: (i) Periodic
orbits exist at rational intrinsic frequencies, similar to the
Poincar\'e-Birkhoff theorem for \twoD{} maps, or (ii) large bends of
the skeleton may occur at resonances involving the normal component. In
combination of these results we generalize the islands-around-islands
hierarchy known from \twoD{} maps to the regular tori in the \fourD{}
phase space, based on the hierarchy of families of $1$-tori.

This paper is organized as follows: In Sec.~\ref{sec:psslcs} we
introduce the map and review the \psslcs{}. In
Sec.~\ref{sec:center-lines} they are used to visualize the $1$- and
$2$-tori and are related to \threeD{} projections. Sec.~\ref{sec:freq}
presents a frequency analysis of the $1$- and $2$-tori. In
Sec.~\ref{sec:explanation} the origin of the second type of families
of $1$-tori is explained and in Sec.~\ref{sec:gaps} the resonance gaps
are discussed. In Sec.~\ref{sec:hierarchy} the hierarchical structure
of the \fourD{} phase space is described. Finally,
Sec.~\ref{sec:conclusion} gives a summary and an outlook. The
algorithm to compute $1$-tori is presented in
Appendix~\ref{sec:computing-central-tori} and a review and
illustration of the break-up of resonant $2$-tori is given in
Appendix~\ref{sec:review-todesco}.

\section{\label{sec:organization}Organization of phase space by 1-tori}

\subsection{\label{sec:psslcs}Coupled standard maps and 2-tori}

\begin{figure*}
  \includegraphics{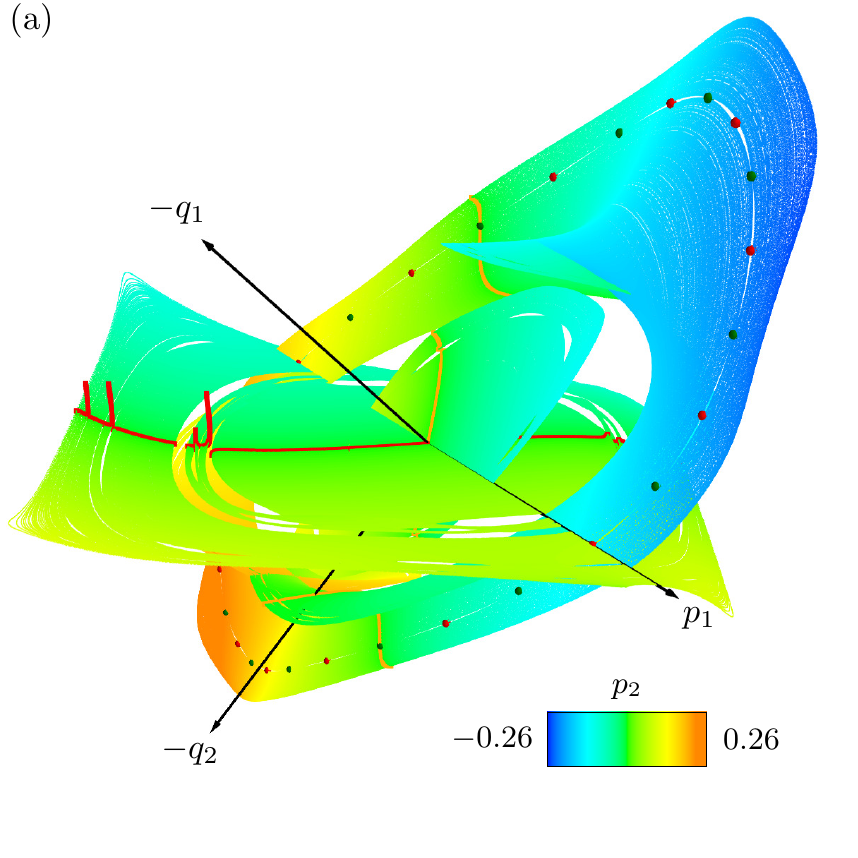}
  \includegraphics{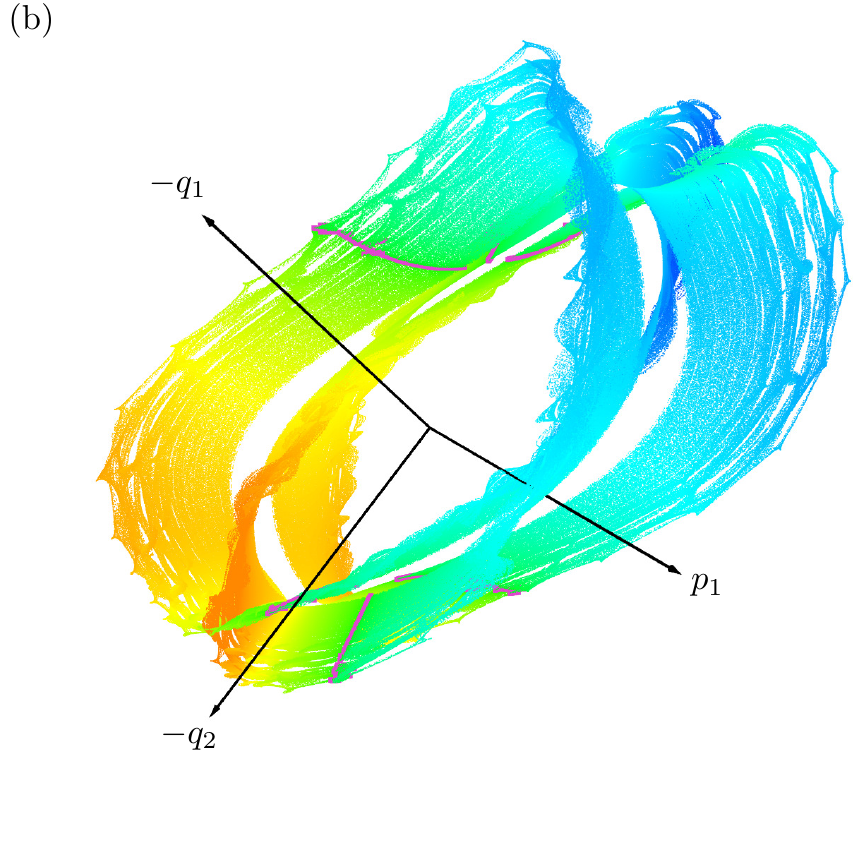}
  \caption{\label{fig:proj-central-tori}%
    Visualization of the \twoD{} Cantor manifolds formed by the
    families of $1$-tori \Ma, \Mb, \Mres{}. For a selection of $1000$
    $1$-tori (excluding the strong bends of \Mb{}) $2000$ iterates
    under the map~(\ref{eq:CoupledStdMaps}) are computed and then
    projected on $(p_1, q_1, q_2)$. The value of $p_2$ is encoded in
    color. The central lines \Ca, \Cb, \Cres, which are the
    intersections of the families of $1$-tori with $p_2=0$, are shown
    in the same colors as in Fig.~\ref{fig:psslc-central-tori}.
    \subfiga{} Families \Ma{} and \Mb{}. The spheres in a gap of \Ma{}
    correspond to two elliptic-elliptic (red) and two
    hyperbolic-elliptic (green) periodic orbits of period $7$.
    \subfigb{} Family \Mres{} corresponding to the $-1:3:0$ resonance.
    \MOVIEREF}
\end{figure*}

As a concrete example to study the organization of phase space in
\fourD{} maps we consider the prototypical system of two coupled
standard maps~\cite{Fro1972} %
\begin{align}
  \label{eq:CoupledStdMaps}
  \begin{aligned}
    \p_1' &= \p_1 + %
    \frac{K_1}{2\pi}\sin\left(2\pi \q'_1\right)  + %
    \frac{\xic}{2\pi}\sin\left(
      2\pi (\q'_1 + \q'_2)
    \right)
    \\
    \p_2' &= \p_2 + %
    \frac{K_2}{2\pi}\sin\left(2\pi \q'_2\right)  + %
    \frac{\xic}{2\pi}\sin\left(
      2\pi (\q'_1 + \q'_2)
    \right)
    \\
    \q_1' &= \q_1 + \p_1
    \\
    \q_2' &= \q_2 + \p_2
  \end{aligned}
\end{align}
where $\p_1, \p_2, \q_1, \q_2 \in [-\nicefrac{1}{2}, \nicefrac{1}{2})$
and periodic boundary conditions are imposed in each coordinate.
The resulting map is symplectic.
The parameters $K_1$ and $K_2$ control the nonlinearity of the
individual \twoD{}
standard maps in $(\p_1, \q_1)$ and $(\p_2, \q_2)$, respectively.
The parameter $\xic$ introduces a coupling between the two
degrees of freedom.
We choose $K_1 = -2.25$, $K_2 = -3.0$ and $\xic = 1.0$, such that the
system is strongly coupled and far from
integrability~\cite{RicLanBaeKet2014}. The origin $\fixedpoint =
(\p_1, \p_2, \q_1, \q_2) = (0, 0, 0, 0)$ is an elliptic-elliptic fixed
point. The eigenvalues $(\lambda_1^\fp, \bar{\lambda}_1^\fp,
\lambda_2^\fp, \bar{\lambda}_2^\fp)$ of the linearized dynamics around
$\fixedpoint$ are $(\exp{(\pm \ui \ 2\pi \nua)}, \exp{( \pm \ui \ 2\pi
  \nub)})$ with $(\nua, \nub) = (0.30632, 0.12173)$ and corresponding
eigenvectors $(\vecfixpt{1}{a}, \vecfixpt{1}{b}), (\vecfixpt{2}{a},
\vecfixpt{2}{b})$.

An orbit started at some initial point in the \fourD{} phase space
leads to a sequence of points $(p_1, p_2, q_1, q_2)$ under the map
\eqref{eq:CoupledStdMaps}. Such an orbit can be visualized by using a
\psslc{} $\Geps$, defined by thickening a \threeD{} hyperplane
\symbolsection{} in the \fourD{} phase
space~\cite{RicLanBaeKet2014}. Explicitly, we consider the slice
defined by
\begin{align}
  \label{eq:slice-condition-in-coordinate}
  \Geps &= \left\{ (\p_1, \p_2, \q_1, \q_2) \; \left|
      \rule{0pt}{2.4ex} \; |\p_2-p_2^*| \le \sectioneps \right. \right\}.
\end{align}
with $p_2^* = 0$ and $\sectioneps = 10^{-4}$ as it provides a good view
of most structures of the map (\ref{eq:CoupledStdMaps}). Whenever a
point of an orbit lies within $\Geps$, the remaining coordinates
$(\p_1, \q_1, \q_2)$ are displayed in a \threeD{} plot. Objects of the
\fourD{} phase space will typically appear in the \psslc{} with a
dimension reduced by one, provided the object intersects with the
slice. Thus, a typical $2$-torus will lead to two or more \oneD{}
lines. A periodic orbit will in general not be visible, unless at
least one of its points lies in the \psslc{}.

In Fig.~\ref{fig:psslc-central-tori} several initial conditions
leading to regular orbits are selected and iterated until for each of
them $4000$ points are contained in the \psslc. They lead to \oneD{}
rings shown in \colortwotorus{} in Fig.~\ref{fig:psslc-central-tori}.
In the center of phase space one has the elliptic-elliptic fixed point
\fixedpoint, which is surrounded by a two-parameter family of
$2$-tori. Further away the $2$-tori form
different clusters, each approximately filling a \fourD{} volume in
the \fourD{} phase space. This regular region is embedded in a chaotic
sea (not shown) formed by chaotic orbits. These structures have
previously been discussed in detail~\cite{RicLanBaeKet2014}.

\subsection{\label{sec:center-lines} Skeleton of 1-tori}

Fig.~\ref{fig:psslc-central-tori} shows that the $2$-tori seem to be
organized in some way even far away from the elliptic-elliptic fixed
point \fixedpoint. We demonstrate in this section that the $2$-tori
are grouped around families of elliptic $1$-tori. Note that in the
following $1$-torus always refers to elliptic $1$-torus.

In order to understand this organization in phase space, we start by
considering a $2$-torus and contract its minor radius to zero using
the algorithm described in Appendix~\ref{sec:computing-central-tori}.
The resulting object is a $1$-torus. If such a $1$-torus intersects
with the \psslc{} it will typically lead to two or more points in the
slice. Performing this contraction procedure for many $2$-tori one
gets the result shown in Fig.~\ref{fig:psslc-central-tori}: The points
of the $1$-tori approximately form lines in the \psslc{}. This
means that the $1$-tori form one-parameter families \M{} in the \fourD{} phase
space. We denote their representation in the \psslc{} as {\it central lines \C{}}. In
Fig.~\ref{fig:psslc-central-tori} all $2$-tori appear to be centered
around one of these lines. This is particularly well visible, e.g.,
for the tori around the center lines \Ca{} (\colorcentera{}) and
\Cres{} (\colorthreetower{}). Thus, the families of $1$-tori \M{} provide the
skeleton for the surrounding $2$-tori. The $1$-tori composing a
particular family \M{} can also be displayed by \threeD{}
projections encoding the value of the projected coordinate by a color
scale~\cite{PatZac1994,KatPat2011}, see
Fig.~\ref{fig:proj-central-tori}.

We now discuss the two types of families of $1$-tori \M{}. This will
be done starting from their representation as central lines \C{} in
the \psslc{}, see Fig.~\ref{fig:psslc-central-tori}:

\alp{} The central lines \Ca{} and \Cb{} shown in \colorcentera{} and
\colorcenterb{} in Fig.~\ref{fig:psslc-central-tori} both emanate from
the central elliptic-elliptic fixed point \fixedpoint{}. Further away
from \fixedpoint{} one has visible gaps but the central lines continue
beyond them. Such gaps result from resonances, see
Secs.~\ref{sec:freq} and \ref{sec:gaps}, and occur on arbitrarily fine
scales along the central lines. Thus more precisely, they are {\it
  Cantor central lines}. In the \threeD{} projection in
Fig.~\ref{fig:proj-central-tori}~\subfiga{} the families of $1$-tori
\Ma{} and \Mb{} compose two 2D Cantor manifolds. These manifolds \Ma{}
and \Mb{} only intersect in the central elliptic-elliptic fixed point
\fixedpoint{}~\cite{Note3}. At \fixedpoint{} the two \twoD{} planes
spanned by the eigenvectors $(\vecfixpt{1}{a}, \vecfixpt{1}{b})$ and
$(\vecfixpt{2}{a}, \vecfixpt{2}{b})$ of the linearized map are
tangential to the manifolds \Ma{} and \Mb{}. Also the frequencies of
the $1$-tori converge to those of \fixedpoint, see
Sec.~\ref{sec:freq}.

These features remind of the invariant manifolds predicted by the
Lyapunov center theorem~\cite{MeyHalOff2009}. However, since the
system is not integrable, \Ma{} and \Mb{} are Cantor manifolds,
interrupted by many gaps due to resonances. They are called
\emph{Lyapunov families of invariant curves}~\cite{JorVil1997b,
  JorOll2004} or \emph{Cantorian central
  submanifolds}~\cite{JorVil1997} and have previously been studied in
the near-integrable regime~\cite{Sev1987,Sev1990,SevLah1991}. Hence,
the central lines \Ca{} and \Cb{} represent the Lyapunov families
\Ma{} and \Mb{} of \fixedpoint{}, respectively. We observe that these
families also exist far away from the elliptic-elliptic fixed point
\fixedpoint{} and continue beyond large resonance gaps.

The central lines \Cpo{} shown in \colorperiodseven{} in
Fig.~\ref{fig:psslc-central-tori} are associated with two
elliptic-elliptic periodic orbits of period $7$. Each torus consists
of seven disjoint parts in the \fourD{} phase
space~\cite{RicLanBaeKet2014}. The central lines emanating from each
point of the elliptic-elliptic period-$7$ orbits are conceptually the
same as the central lines of the central elliptic-elliptic fixed point
\fixedpoint. In Appendix~\ref{sec:review-todesco} it is explained that
the period-$7$ orbits exist due to a rank-$2$ resonance.

\bet{} The central lines \Cres{} shown in \colorthreetower{} in
Fig.~\ref{fig:psslc-central-tori} are fundamentally different from
\Ca, \Cb, and \Cpo. They occur as two groups of three branches, each
group emerging near the central line \Ca{}
(\colorcentera). Most importantly, there is no fixed point or periodic
orbit from which the central lines \Cres{} originate. The
visualization of the family of $1$-tori \Mres{} in
Fig.~\ref{fig:proj-central-tori}~\subfigb{} has a very different
topology from case \alp{} in
Fig.~\ref{fig:proj-central-tori}~\subfiga{}. It is shown in
Sec.~\ref{sec:explanation}, that such types of families of $1$-tori
result from broken resonant $2$-tori of a rank-$1$ resonance. In the
case of \Cres{} it is the $-1:3:0$ resonance. Still, we make the same
observations regarding resonance gaps as in case \alp{}, see
Sec.~\ref{sec:gaps}.

Note that there is a case where the types \alp{} and \bet{}
coincide. Such families of $1$-tori are discussed in
Sec.~\ref{sec:hierarchy}.

\begin{figure}[b]
  \includegraphics{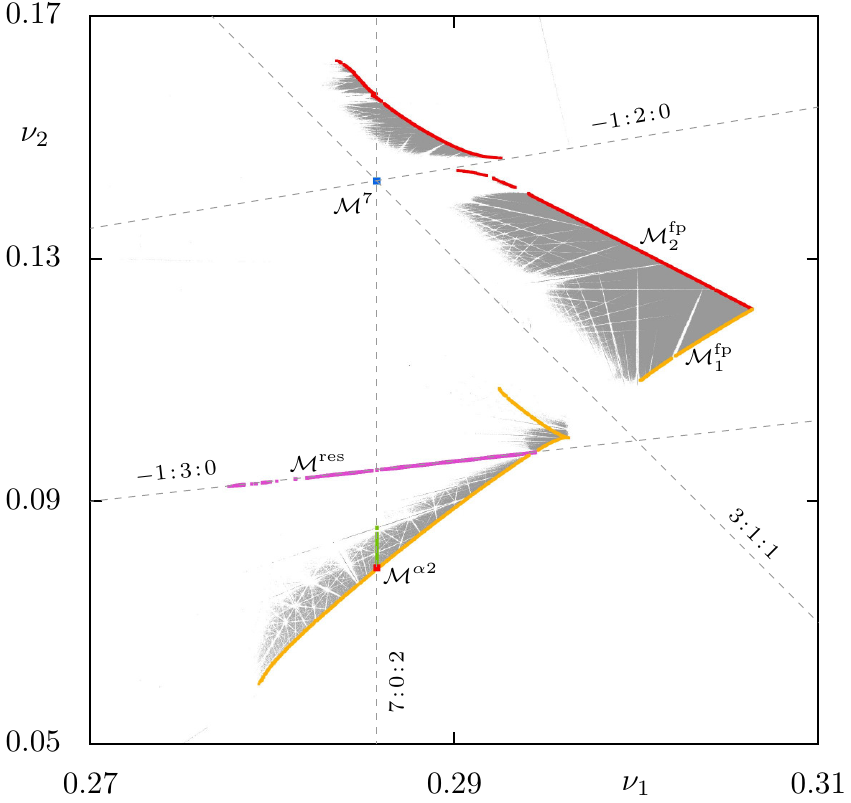}
  \caption{\label{fig:fa-central-tori}%
    Frequencies $(\nu_1, \nu_2)$ of all $2$-tori (grey). The largest
    gaps stem from the $3:1:1$ and $-1:2:0$ resonances. The
    frequencies of the fixed point \fixedpoint{} are $(\valfixpt{1},
    \valfixpt{2})=(0.30632, 0.12173)$. From there the \alp{} families
    of $1$-tori \Ma, \Mb{} (\colorcentera{}, \colorcenterb{}) emanate,
    forming the skeleton of the regular region. The frequencies for
    $1$-tori of the \bet{} family \Mres{}
    (\colorresonancethreetower{}) are all on the $-1:3:0$ resonance
    line and for the \alp{} families \Mpo{} (\colorperiodseven{}) on
    the point $(\nicefrac{2}{7}, \nicefrac{1}{7})$. The families
    \Mh{\alpm{2}}{} (\coloralphab) are discussed in Sec.~\ref{sec:hierarchy}.}
\end{figure}

\subsection{\label{sec:freq}1-tori in the frequency plane}

In order to understand the influence of resonances we now relate the
observations in phase space to a frequency
analysis\cite{Las1993,BarBazGioScaTod1996} which associates with each
$2$-torus its two fundamental frequencies $(\nu_1, \nu_2) \in
[0,1[^2$. These are displayed in the frequency plane, see
Fig.~\ref{fig:fa-central-tori}, where the \colortwotorus{} points
represent $2$-tori obtained by starting $10^8$ initial conditions with
randomly chosen $p_1, p_2,q_1, q_2 \in [-0.2, 0.2]$ in the \fourD{}
phase space. Each frequency pair is calculated from $N=4096$
iterations. To decide whether an orbit is regular we use the frequency
criterion
\begin{align}
  \label{eq:frequency-criterion}
  \max{\left(|\nu_1 - \tilde{\nu}_1|,|\nu_2 - \tilde{\nu}_2|\right)} < 10^{-7},
\end{align}
where the frequency pair $(\tilde{\nu}_1, \tilde{\nu}_2)$ is
calculated from $N$ further iterations. This leads to nearly $3 \cdot
10^6$ regular $2$-tori. Since the frequencies $(\nu_1, \nu_2)$ are
only defined up to a unimodular
transformation~\cite{Bor1927,DulMei2003,GekMaiBarUze2007} we use the
\psslcs{} to choose frequencies consistently~\cite{RicLanBaeKet2014}.
The frequency plane is covered by resonance lines $m_1:m_2:n$, on
which the frequencies fulfill
\begin{align}
  \label{eq:resonance-driven}
  m_1 \cdot \nu_1 + m_2 \cdot \nu_2 = n
\end{align}
where $m_1, m_2, n$ are integers with at least one being nonzero. The
most important resonance lines are displayed in
Fig.~\ref{fig:fa-central-tori}.

The two-parameter families of $2$-tori in the \psslc{} directly
correspond to areas in the frequency plane~\cite{RicLanBaeKet2014}.
For example, the frequencies (\nua, \nub) of the elliptic-elliptic
fixed point \fixedpoint{} correspond to the rightmost tip. Large gaps
due to resonances, like the $3:1:1$ and the $-1:2:0$, correspond to
gaps in the \psslc{}, see Fig.~\ref{fig:psslc-central-tori}.

In contrast to a $2$-torus, the dynamics on a $1$-torus is described
by only one frequency, which is also called intrinsic
frequency. For example, the frequency on \Ma{}
(\colorcentera{}) corresponds to $\nu_1$ and on \Mb{}
(\colorcenterb{}) to $\nu_2$ in Fig.~\ref{fig:fa-central-tori}. The
frequency describing the normal behavior can be obtained from the
limiting $2$-tori or from the linearized dynamics normal to the
$1$-torus~\cite{Jor2001,JorOll2004}. This frequency is called normal
or librating frequency. The families of $1$-tori are indicated in
Fig.~\ref{fig:fa-central-tori} with colors corresponding to their
central lines in Fig.~\ref{fig:psslc-central-tori}. This allows to
make the connection with the two types \alp{} and \bet{} of families
of $1$-tori \M{} represented by the central lines \C{}:

\alp{} The sharp edges emanating from the elliptic-elliptic fixed
point \fixedpoint{} at (\nua, \nub) correspond to the families \Ma{}
and \Mb{} (\colorcentera{}, \colorcenterb{}). The frequencies of the
$2$-tori emanate from these edges. In this sense, the $1$-tori also
organize the frequencies of the $2$-tori. Moreover, one recognizes in
the frequency plane that the gaps in the families \M{} and their
central lines \C{} in Figs.~\ref{fig:psslc-central-tori}
and~\ref{fig:proj-central-tori} are due to resonances.
Fig.~\ref{fig:fa-central-tori} also nicely illustrates that the edges
corresponding to \Ma{} and \Mb{} continue beyond large resonance gaps.
Note that the frequencies of the $1$-tori of \Mpo{} and their elliptic
surrounding collapse to the point $(\nu_1, \nu_2) = (\nicefrac{2}{7},
\nicefrac{1}{7})$.

\bet{} The frequencies of the family of $1$-tori \Mres{}
(\colorthreetower{}) lie on the $-1:3:0$ resonance line in
Fig.~\ref{fig:fa-central-tori}. The frequencies of their surrounding
$2$-tori also lie on this resonance line.

The reason that the families \Mpo{} and \Mres{} and their surrounding
$2$-tori collapse on a point or a line, respectively, in
Fig.~\ref{fig:fa-central-tori} is that here the frequencies are
calculated with respect to the central elliptic-elliptic fixed point
\fixedpoint{}. In Sec.~\ref{sec:hierarchy} an adapted frequency
analysis is performed for these cases.

\subsection{Families of 1-tori from rank-1 resonances}
\label{sec:explanation}

We now discuss that families of $1$-tori of type \bet{}, such as \Mres,
which do not correspond to Lyapunov families of elliptic-elliptic
fixed points or periodic orbits, originate from broken $2$-tori that
fulfilled a rank-$1$ resonance condition. This follows by applying
results of Todesco~\cite{Tod1994}, which are obtained for the vicinity
of an elliptic-elliptic fixed point, to arbitrary families of resonant
$2$-tori far away from a fixed point or even in the absence of a
fixed point. This is reviewed and illustrated in
Appendix~\ref{sec:review-todesco}: One of the results is that a
$2$-torus fulfilling a rank-$1$ resonance in an integrable \fourD{}
symplectic map breaks up into several $1$-tori when a perturbation is
added. More precisely, an equal number of elliptic and hyperbolic
$1$-tori remains after the break-up of the $2$-torus. Since for each
rank-$1$ resonance there is usually a one-parameter family of $2$-tori
in the integrable system we infer that in the perturbed system for
each of these resonances a one-parameter Cantor family of elliptic
$1$-tori exists. The appearance of these families in phase space is
governed by the position and dynamics of the original resonant
$2$-tori, see Appendix~\ref{sec:review-todesco}.

The $1$-tori of the central lines \Cres{} (\colorthreetower{}) in
Fig.~\ref{fig:psslc-central-tori} originate from the
one-parameter family of broken $2$-tori whose frequencies lie on the $-1:3:0$
resonance line. This is in accordance with the frequencies of the
$1$-tori of the family \Mres{} in Fig.~\ref{fig:fa-central-tori}. Note
that we only show the elliptic and not the hyperbolic $1$-tori of the
broken $2$-tori in Figs.~\ref{fig:psslc-central-tori} and
~\ref{fig:proj-central-tori}~\subfigb{}. While \Mres{} is an example
for a coupled rank-$1$ resonance, we make analogous observations for
uncoupled rank-$1$ resonances, e.g., for the $7:0:2$ resonance in
Appendix~\ref{sec:review-todesco}.

\subsection{Resonance gaps in the skeleton of 1-tori}
\label{sec:gaps}

\begin{figure*}
  \includegraphics{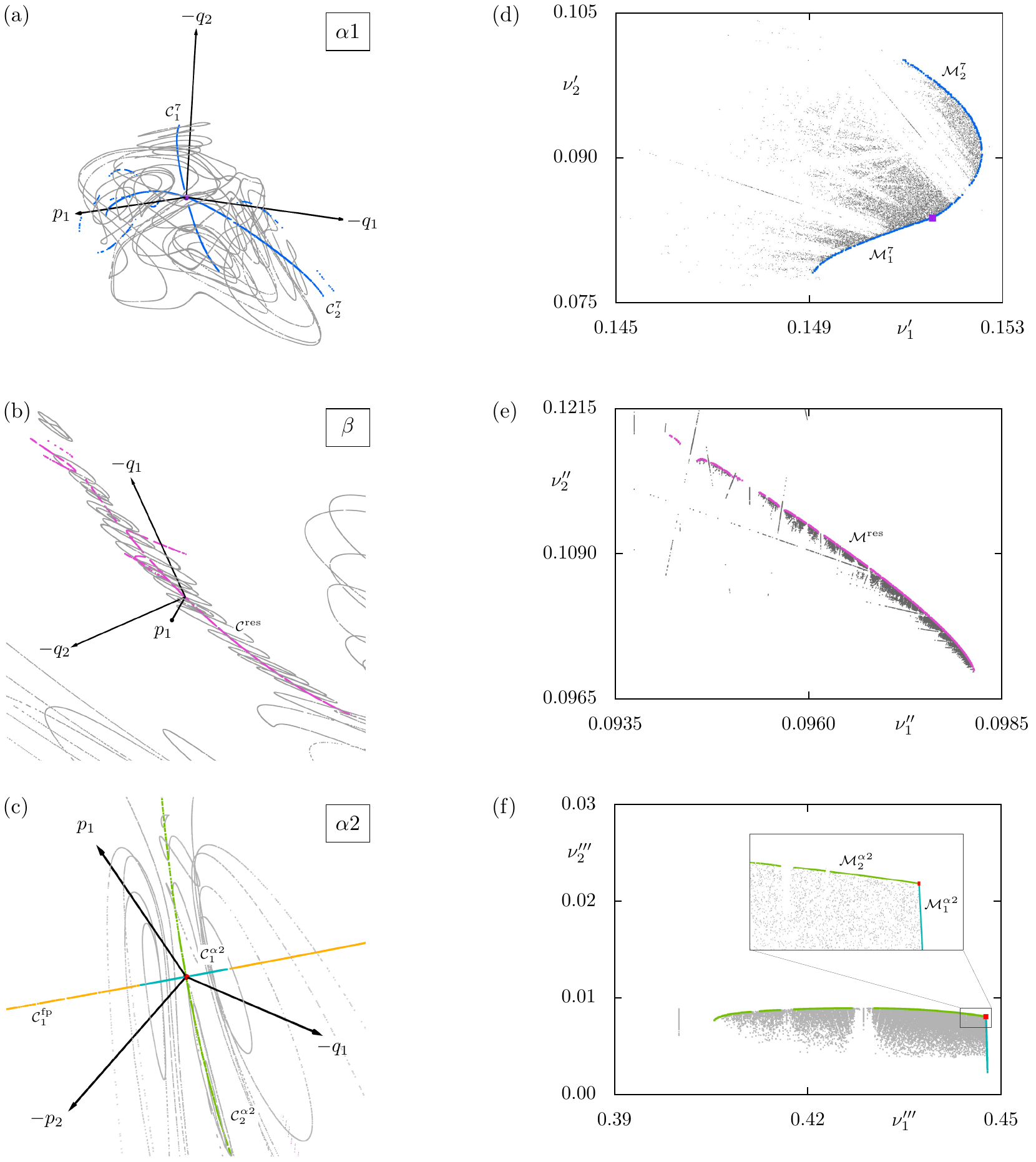}
  \caption{\label{fig:hierarchy}%
    First level of the hierarchy for \subfiga{} type \alp{1},
    \subfigb{} type \bet{}, and \subfigc{} type \alp{2}: Magnified
    views of Fig.~\ref{fig:psslc-central-tori} for \subfiga{} \Cpo{}
    around one point $\periodicpoint=(0.0, 0.0, 0.083438087,
    0.118666288)$ of the elliptic-elliptic period-$7$ orbit,
    \subfigb{} a branch of \Cres{} of the $-1:3:0$ resonance, and
    \subfigc{} central lines \Caa{1} (\coloralphaa) and \Caa{2}
    (\coloralphab) representing families of $1$-tori \Maa{1} and
    \Maa{2} around one point $\alphapoint=(0.115287658, -0.141621338,
    0.0, 0.0)$ of the elliptic-elliptic period-$7$ chain. The slice
    condition is for \subfiga{}, \subfigb{} $p_2^*=0$ and
    $\sectioneps=10^{-5}$ and for \subfigc{} $q_2^*=0$ and
    $\sectioneps=10^{-6}$. The coordinate system in \subfiga{} and
    \subfigc{} is moved to \periodicpoint{} and \alphapoint{}
    respectively, and in \subfigb{} to an arbitrary point along \Cres.
    In \subfigd{}, \subfige{}, and \subfigf{} the adapted frequencies
    are shown. In \subfiga{} and \subfigd{} the periodic point
    \periodicpoint{} with frequencies $(\nu_1', \nu_2')=(0.1515,
    0.0838)$ is marked by a magenta point. The two emanating families
    of $1$-tori are colored the same for simplicity and have edges
    intersecting in the frequency plane with an angle close to $\pi$.
    In \subfigc{} and \subfigf{} the periodic point \alphapoint{} with
    frequencies $(\nu_1''', \nu_2''')=(0.4475, 0.0081)$ is marked by a
    red point. The inset in \subfigf{} shows a magnification around
    \alphapoint. \MOVIEREF}
\end{figure*}

As mentioned in the previous sections the skeleton of $1$-tori is
interrupted by gaps due to resonances. Two different types of such
gaps are observed depending on whether the normal frequency is
involved in the resonance:

(i) If the intrinsic frequency along a family of $1$-tori crosses a
rational number, we observe a chain of alternating elliptic-elliptic
and elliptic-hyperbolic periodic orbits arranged on a \oneD{} line.
This reminds of a break-up of a resonant $1$-torus according to the
Poincar\'e-Birkhoff theorem in \twoD{} maps. As an example, we show in
Fig.~\ref{fig:proj-central-tori}~\subfiga{} two elliptic-elliptic and
two elliptic-hyperbolic periodic orbits of period $7$, forming a chain
in the gap of the family \Ma{}. As seen in
Fig.~\ref{fig:fa-central-tori}, these period-$7$ orbits arise from the
intersection (red) of the $7:0:2$ resonance (\coloralphab) with the
family \Ma{} (\colorcentera). The
emanating families of $1$-tori \Maa{} and their surrounding are
discussed in Sec.~\ref{sec:hierarchy} and shown in
Fig.~\ref{fig:hierarchy}~\subfigc{}.

(ii) If the frequencies along a family of $1$-tori cross a resonance
involving the normal frequency, the central lines in the vicinity of
the resonance may bend on either side of
the resonance gap. This is well visible in
Fig.~\ref{fig:psslc-central-tori} for the big gap of the central line
\Ca{} (\colorcentera) caused by the $3:1:1$ resonance and the gap of
the central line \Cb{} (\colorcenterb) caused by the $-1:2:0$
resonance. This behavior coincides with a bending of the corresponding
edges in the frequency plane, see Fig.~\ref{fig:fa-central-tori}. For
a system which is far from being integrable this bending in phase
space and the frequency plane can occur on large scales and can have
substantial impact on the dynamics. For instance, the regular region
around the larger bend of \Ca{} at the $3:1:1$ resonance is the main
region for power-law trapping in the system~\cite{Lan2012}.

Note that for the $-1:3:0$ resonance the corresponding central line
\Cres{} does not connect with the central line \Ca{} in
Fig.~\ref{fig:psslc-central-tori}. This is similar to the period
tripling bifurcation for area-preserving \twoD{} maps.

These observations correspond well to results about bifurcations of a
$1$-torus in volume-preserving \threeD{}
diffeomorphisms~\cite{DulMei2009} and in a quasi-periodically forced
Hamiltonian oscillator~\cite{BroHanJorVilWag2003}. In particular,
there it is observed that bends occur at resonances
$m_{\text{int}}:m_{\text{norm}}:n$ at which the coefficient
$m_{\text{norm}}$ of the normal frequency fulfills
$|m_{\text{norm}}|=1$. This condition is fulfilled for \Ma{} crossing
the $3:1:1$ resonance and \Mb{} crossing the $-1:2:0$ resonance. A
more detailed investigation of the resonance gaps, which also takes
into account hyperbolic $1$-tori, is left for the future.

\section{Hierarchy of regular tori}
\label{sec:hierarchy}

It is well-established that the phase space of \twoD{} maps
generically exhibits a hierarchy~\cite{LicLie1992,Mei1986} of regular
tori, i.e., \emph{islands-around-islands} and island chains converging
towards an irrational torus. In this case the regular tori are organized
around elliptic periodic orbits. We now discuss for \fourD{} maps a
hierarchy similar to the islands-around-islands case. Since $2$-tori are
organized around elliptic $1$-tori, it is sufficient to understand the
hierarchy of families of elliptic $1$-tori. In this sense, these
families of elliptic $1$-tori correspond to the elliptic periodic
orbits in \twoD{} maps.

According to the previous section there are two possible origins for
families of elliptic $1$-tori: \alp{} A family emanates from an
elliptic-elliptic point or \bet{} it results from broken resonant
$2$-tori that fulfilled a rank-$1$ resonance. The elliptic-elliptic
points in case \alp{} can be either due to \alp{1} a broken $2$-torus
that fulfilled a rank-$2$ resonance, see
Appendix~\ref{sec:review-todesco}, or \alp{2} a broken elliptic
$1$-torus with rational intrinsic frequency $\nu_1=m/n$, i.e., a
rank-$1$ resonance $m:0:n$, see Sec.~\ref{sec:gaps}~(i).

There is a hierarchical structure, where families of elliptic $1$-tori
are organized around a family of elliptic $1$-tori. Consider such a
family of elliptic $1$-tori \Mh{}{} with intrinsic frequencies
$\nu_1$, which is surrounded by its two-parameter family of $2$-tori
\Th{}{}. The next level of the hierarchy around \Mh{}{} results from
three types of structures:

\alp{1} rank-$2$ resonances in \Th{}{}, each giving elliptic-elliptic
points with two families \Mh{\alpm{1}}{1} and \Mh{\alpm{1}}{2},

\alp{2} rank-$1$ resonances $m:0:n$ on \Mh{}{}, i.e., rational
intrinsic frequencies, each giving elliptic-elliptic points with two
families \Mh{\alpm{2}}{1} and \Mh{\alpm{2}}{2},

\bet{} rank-$1$ resonances in \Th{}{}, each giving a family
\Mh{\betm{}}{}.

Around each of these new families of $1$-tori the hierarchy continues
with these three types \alp{1}, \alp{2} and \bet{} on finer and finer
scales. The case \alp{1} might be seen as a direct generalization of
the islands-around-islands hierarchy in \twoD{}
maps~\cite{LicLie1992,Mei1986}. It is interesting to observe that in
case \alp{2} the family \Mh{\alpm{2}}{1} is tangential to the family
\Mh{}{}, looking like an island chain within \Mh{}{}. The other family
\Mh{\alpm{2}}{2} coincides with the type \bet{} family of the rank-$1$
resonance $m:0:n$.

We now explicitly illustrate the hierarchy of $1$-tori starting from the
elliptic-elliptic fixed point \fixedpoint{} with its two Lyapunov
families \Ma{} and \Mb{} of $1$-tori. For
all three types of structures Figs.~\ref{fig:hierarchy}~\subfiga{},
\subfigb{}, and \subfigc{} show $2$-tori around \Mpo{}, \Mres{}, and
\Maa{1,2} in the \psslc{} and Figs.~\ref{fig:hierarchy}~\subfigd{},
\subfige{}, and \subfigf{} show the corresponding frequency planes. For
this we calculate the frequencies of each torus with respect to its
organizing structure (a similar method has been used for time-continuous
systems~\cite{PapLas1998}). That is, for a torus around \Mpo{} or
\Maa{1,2} only every $7$-th point is used to obtain the frequencies
$(\nu_1',\nu_2')$ and $(\nu_1''',\nu_2''')$ respectively. For a torus
around \Mres{} additionally to the frequency $\nu_1'' = \nu_1/3 =
\nu_2$ another independent frequency $\nu_2''$ is computed.
Fig.~\ref{fig:hierarchy} provides examples for all three cases:

\alp{1} The families \Mpo{1} and \Mpo{2} around one of the
elliptic-elliptic period-$7$ orbits lead to \Cpo{1} and \Cpo{2} in
Fig.~\ref{fig:hierarchy}~\subfiga{}. The regular regions around such
an elliptic-elliptic point \periodicpoint{} and the elliptic-elliptic
fixed point \fixedpoint{}, see Fig.~\ref{fig:psslc-central-tori}, look
qualitatively similar. The same is true for the frequency planes in
Figs.~\ref{fig:hierarchy}~\subfigd{} and \ref{fig:fa-central-tori}.

\bet{} \Mres{}, which originates from the $-1:3:0$ resonance, leads to
\Cres{} in Fig.~\ref{fig:hierarchy}~\subfigb{}. The regular regions for such a
resonance in Fig.~\ref{fig:hierarchy}~\subfigb{}
and~\ref{fig:hierarchy}~\subfige{} are different from those shown in
Figs.~\ref{fig:psslc-central-tori}, \ref{fig:fa-central-tori},
\ref{fig:hierarchy}~\subfiga{}, and \ref{fig:hierarchy}~\subfigd{} as
there is only one family of $1$-tori and only one edge in the
frequency plane.

\alp{2} \Maa{1} and \Maa{2} resulted from the intersection of the
$7:0:2$ resonance with \Ma{} and lead to \Caa{1} and \Caa{2} in
Fig.~\ref{fig:hierarchy}~\subfigc{}. The regular region around \Maa{1}
and \Maa{2} in Figs.~\ref{fig:hierarchy}~\subfigc{} and \subfigf{}
looks qualitatively similar to the case \alp{1}. However, one family
\Maa{1} leading to the central line \Caa{1} (\coloralphaa) is embedded
in a gap of \Ca{} (\colorcentera) in
Fig.~\ref{fig:hierarchy}~\subfigc{}. Consequently, \Caa{1} is quite
short. \Maa{2}, leading to \Caa{2} (\coloralphab) in
Fig.~\ref{fig:hierarchy}~\subfigc{}, coincides with the type \bet{}
family originating from the $7:0:2$ resonance, see
Fig.~\ref{fig:fa-central-tori}. As none of the points of the
elliptic-elliptic period-$7$ orbit lie in the $p_2^*=0$ slice, see
Fig.~\ref{fig:proj-central-tori}~\subfiga{}, they are not visible in
Fig.~\ref{fig:psslc-central-tori}. Thus, we use in
Fig.~\ref{fig:hierarchy}~\subfigc{} the slice condition $q_2^*=0$.

Note that, as in the case of \twoD{} maps, there may be periodic
orbits that neither correspond to case \alp{1} nor \alp{2}. They
result from saddle-node bifurcations~\cite{KooMei1989a} and generate
their own hierarchy in the above sense. Such bifurcations have to be
investigated with parameter studies~\cite{MaoSatHu1985}, which are not
performed here.

\section{\label{sec:conclusion}Summary and outlook}

In this paper we study the organization of regular tori in a generic
\fourD{} map. Using \psslcs{} we visualize how the $2$-tori are
arranged around a skeleton of elliptic $1$-tori, which are computed
using a new iterative contraction method. This provides a
generalization of the well-known case of \twoD{} maps, where regular
tori encircle elliptic fixed points or periodic orbits.

The $1$-tori occur in one-parameter families, appearing as Cantor
central lines in the \psslc. Two types of families can be
distinguished: \alp{} Families emanating from an elliptic-elliptic fixed
point or periodic orbit. While this type is known from near-integrable
systems to correspond to Lyapunov families of invariant curves, we
observe that these families exist far away from the elliptic-elliptic
points and continue beyond large resonance gaps. We show that type \bet{}
results from broken resonant $2$-tori that fulfilled a rank-$1$
resonance condition.

The skeleton of $1$-tori is interrupted by resonance gaps. We observe
two distinct behaviors when a family of $1$-tori crosses a resonance:
(i) If the resonance involves only the intrinsic frequency of the
$1$-tori, periodic orbits occur similar to the case of the
Poincar\'e-Birkhoff theorem of \twoD{} maps. (ii) If the resonance
involves the normal frequency of the $1$-tori, the family may bend on
either side of the resonance gap. At low order resonances these bends
have substantial impact on the geometry and dynamics in phase space.
The two types of families of $1$-tori, their interplay with
resonances, and the break-up of resonant $2$-tori allow for
interpreting all observed regular structures. These results also
explain the hierarchy of the regular structures and thus
generalize the well-known islands-around-islands hierarchy of \twoD{}
maps.

As an outlook we mention that it should be possible to extent the
results to understand the
origin of lower-dimensional tori and their hierarchy in even
higher-dimensional systems. Moreover, the visualization of central
lines formed by elliptic $1$-tori should be complemented by a
computation of hyperbolic $1$-tori. Another line of investigation is
to have a more detailed look at the resonance gaps and bifurcations of
the families of $1$-tori.

We believe that the computation and visualization of the skeleton of
$1$-tori can also be useful for other systems like the helium atom,
the hydrogen atom in crossed fields, or the restricted three-body
problem to understand the organization of regular tori in phase space.

\begin{acknowledgments}
  We are grateful for discussions with Jacques Laskar, Jim Meiss,
  Haris Skokos and Holger Waalkens. We would like to thank the
  referees for very useful comments. Furthermore, we acknowledge
  support by the Deutsche Forschungsgemeinschaft within the
  Forschergruppe 760 ``Scattering Systems with Complex Dynamics.'' All
  \threeD{} visualizations were created using
  \mayavi\cite{RamVar2011}.
\end{acknowledgments}

\appendix

\section{Computation of elliptic 1-tori}
\label{sec:computing-central-tori}

While $f$-tori in a $2f$-dimensional phase space can be found by a
brute-force search and many algorithms exist to determine periodic
orbits, lower-dimensional tori have to be obtained from more
sophisticated
algorithms~\cite{Sim1998,CasJor2000,JorOll2004,SchOsiVog2005,
  LanChaCvi2006,HarLla2007,DulMei2009,CouLasCorMayUdr2010,SanLocGio2011,HugLlaSir2012}.
We present a new approach which is based on an iterative contraction
of $f$-tori to lower-dimensional elliptic tori~\cite{Lan2012,Onk2012}.
One of the advantages of this approach is that it has not to be
adjusted for more complicated topologies of $1$-tori, i.e., occurring
at the resonance gaps, and that it does not require any slices.
Moreover, as no continuation is necessary to find families of $1$-tori
the algorithm is not restricted by large resonance gaps. The basic
idea should work for arbitrary dimensions, but for simplicity we only
discuss the case $f=2$.

The idea of the iterative method is to start from a $2$-torus and
geometrically contract it to a $1$-torus, i.e., a $2$-torus for which
one action is $0$: Consider an initial $2$-torus $T^0$ with actions
$(I_1^0, I_2^0)$, corresponding angle coordinates $(\Theta_1^0,
\Theta_2^0)$, and frequencies $(\nu_1^0, \nu_2^0)$. Without loss of
generality, we assume that geometrically $\Theta_1^0$ belongs to the
major radius of the torus. Starting from an initial point on the torus
we obtain an orbit $\vec{x}^0(t)$ and the major frequency $\nu_1^0$.
We determine the times $t_l < t_{\text{max}}$, with $l =
0,1,\ldots,L-1$, for which the angle coordinates $\Theta_1^0(t)$ of
$\vec{x}^0(t)$ are closest to $\Theta_1^0(0)$, i.e.,
\begin{align}
  \Theta_1^0(t_l)-\Theta_1^0(0) &= 2\pi\nu_1^0t_l \ \text{mod} \
  2\pi \approx 0 \ .
\end{align}
The points $\vec{x}^0(t_l)$ differ in their normal angle $\Theta_2^0$
but approximately match in the angle $\Theta_1^0$. Note that a similar
setting is used in time-continuous systems to average out the resonant
motion~\cite{PapLas1998}. One could simply use the geometric center of
the $L$ points $\vec{x}^0(t_l)$ as a new initial point
$\vec{x}^1(0)$~\cite{Lan2012,Onk2012}. Motivated by ellipsoidal
approximations~\cite{DulMei2009}, a faster convergence is achieved by
fitting \twoD{} ellipses~\cite{FitPilFis1999} to the projections of
the points $\vec{x}^0(t_l)$ in each degree of freedom. Their centers
define the new initial point $\vec{x}^1(0)$. The orbit $\vec{x}^1(t)$
for this point lies supposedly on a similar $2$-torus $T^1$ with
smaller action $I_2^1 < I_2^0$. Iterating these steps $N$ times gives
a sequence of $2$-tori $T^i$ approaching a $1$-torus. By construction
the frequency $\nu_1^i$ and action $I_1^i$ of the $2$-tori $T^i$
converge to those of the $1$-torus.

A problem common to all methods computing lower-dimensional tori is to
decide whether the result is close enough to a $1$-torus, ideally by
estimating the minor radius. For this, we try to approximate the orbit
$\vec{x}^N(t)$ on the final torus $T^N$ by a Fourier
series~\cite{BazBonTur1998, CasJor2000, LanChaCvi2006}
\begin{align}
  \label{eq:fourier-series1d}
  \vec{x}(t) &= \sum_{|k| \leq k_{\text{max}}} \vec{c}_k \ue^{\ui
    2\pi k \nu_1^N t} ,
\end{align}
for $t<\tau$. If the maximal square deviation $\sigma_\tau^2 =
\sup_{t<\tau} ||\vec{x}(t) - \vec{x}^N(t)||^2$ of the Fourier series
is smaller than a prescribed value $\sigma_{\text{max}}^2$, the torus
$T^N$ is considered to be a $1$-torus. However, even if this criterion
is fulfilled for a chosen $\tau$ the actual torus may still have a large minor radius. This
becomes only apparent when using more iterations $t>\tau$. Such a
torus usually has frequencies close to a rank-$1$ resonance. Thus,
additionally we use the maximal square deviation $\sigma_{4 \tau}^2 =
\sup_{t<4 \tau} ||\vec{x}(t) - \vec{x}^N(t)||^2$ with the previous
Fourier series $\vec{x}(t)$ obtained from
$t<\tau$. A torus is discarded if the ratio $R_{\sigma^2} = \sigma_{4
  \tau}^2 / \sigma_\tau^2$ is bigger than $R_{\text{max}}$. In such
cases one could also increase $t_{\text{max}}$ to obtain a better
converged torus.

To obtain the $1$-tori used in this paper we start from approximately
$3 \cdot 10^{6}$ $2$-tori, see Sec.~\ref{sec:freq}, and for each of
them use initially $N=10$, $L=10$, $t_{\text{max}} = 2^{14}$,
$\tau=4096$, and the smallest $k_{\text{max}} \in [40, 80, 160, 320,
400]$ for which $\sigma_\tau^2 < \sigma_{\text{max}}^2=10^{-10}$ and
$R_{\sigma^2}<R_{\text{max}}=2.25$ is fulfilled. If these criteria
fail the computation is repeated one more time with doubled
$t_{\text{max}}$. Since the algorithm sometimes converges to periodic
orbits, also $1$-tori with near-rational major frequency $\nu_1^N$ are
excluded. That is, if the highest convergent $a/b$, $a,b \in \N$ of a
continued fraction expansion of $\nu_1^N$ with $b<500$ (or $b<3500$
for the period-$7$ island) fulfills $|\nu_1^N-a/b|<10^{-10}$ it is
discarded. Finally, the intersections of converged tori with the
\psslc{} are obtained by linear interpolation using $10^4$ iterates of
each torus. The result is accepted if the points used for the
interpolation are closer to the slice than $10^{-3}$. Larger distances
result from a near-rational major frequency and could be treated by
using more iterates or higher order interpolation methods. Some
converged tori intersect the slice more often due to strong bends or a
shallow intersection angle. To display only the main skeleton of
$1$-tori, we excluded such tori if in case of \Ma{} and \Mb{}, any of
their intersections with the slice does not fulfill $|p_1|<10^{-3}$
and, in case of \Mres{}, none of the intersections fulfills
$|p_1|<10^{-3}$.

An advantage of the algorithm is that only the major frequency $\nu_1$
has to be computed correctly to obtain the right central lines: For
example to get the Lyapunov families of the elliptic-elliptic fixed
point \fixedpoint{} the frequency with largest amplitude is used. For
the Lyapunov families of the elliptic-elliptic period-$7$ orbits the
frequency with largest amplitude with respect to the $7$-th iterate of
the map is chosen. For the $-1:3:0$ resonance the frequency $\nu_1/3 =
\nu_2$ is used. Note that the algorithm can also converge to $1$-tori
of rank-$1$ resonances $m_1:m_2:n$ at which the coefficient $m_2$ of
the normal frequency, or in this context minor frequency, fulfills
$|m_2|=1$, e.g., the $7:1:2$ resonance at \Ma{}. These $1$-tori have
been omitted in the figures for simplicity.

\section{Break-up of resonant 2-tori}
\label{sec:review-todesco}

In this section, we briefly review and illustrate the results of
Todesco about the break-up of resonant $2$-tori of an integrable
\fourD{} symplectic map when a perturbation is added~\cite{Tod1994}.
These results are derived for the vicinity of an elliptic-elliptic
fixed point, but we find the described behavior for resonant tori far
away from a fixed point or even in the absence of a fixed point. Also, the examples
demonstrate how the geometry of the original resonant $2$-torus
governs the geometry of the phase-space structures remaining after the
break-up. Additionally, the break-up of a $1$-torus with rational
intrinsic frequency is presented, a case not discussed by
Todesco.

\begin{figure*}
  \includegraphics{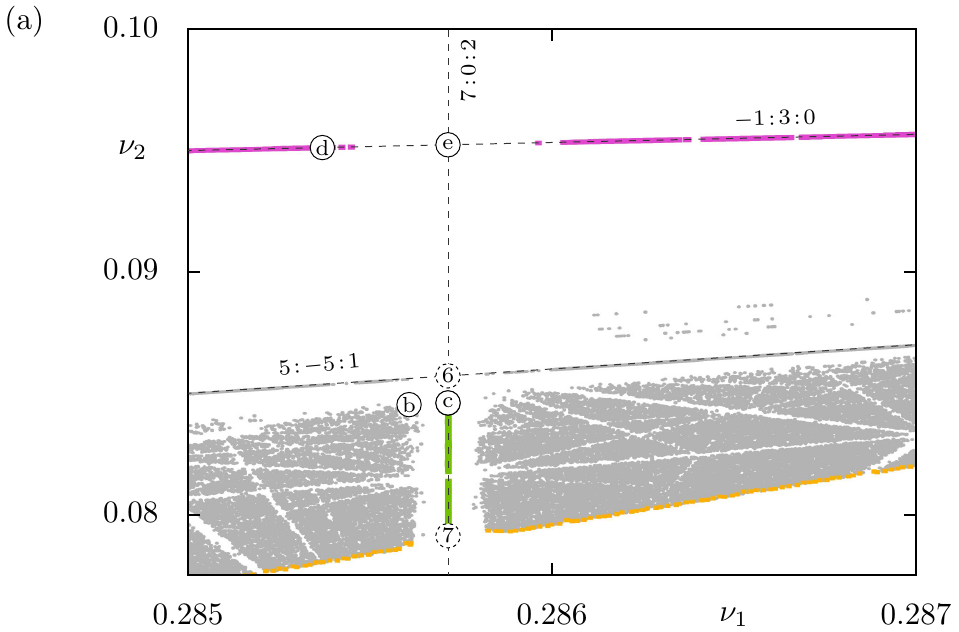}\\
  \vspace*{0.5cm}
  \includegraphics{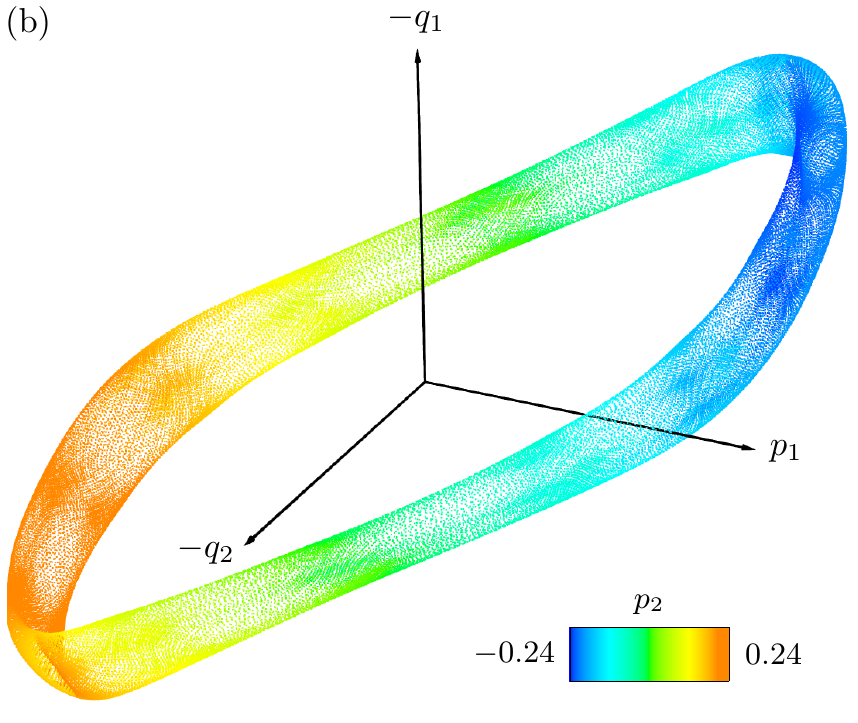}
  \includegraphics{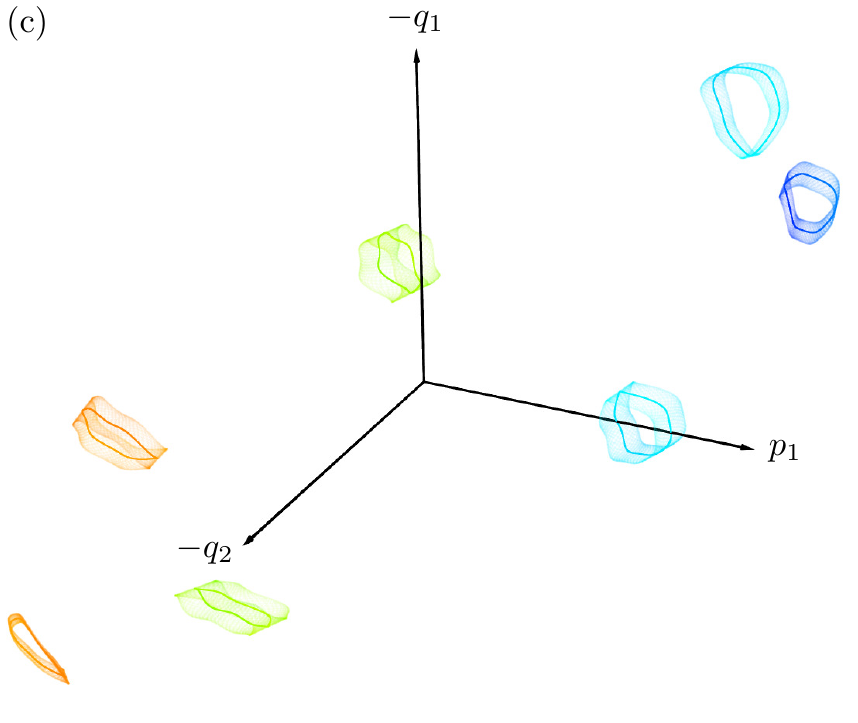}\\[2ex]
  \includegraphics{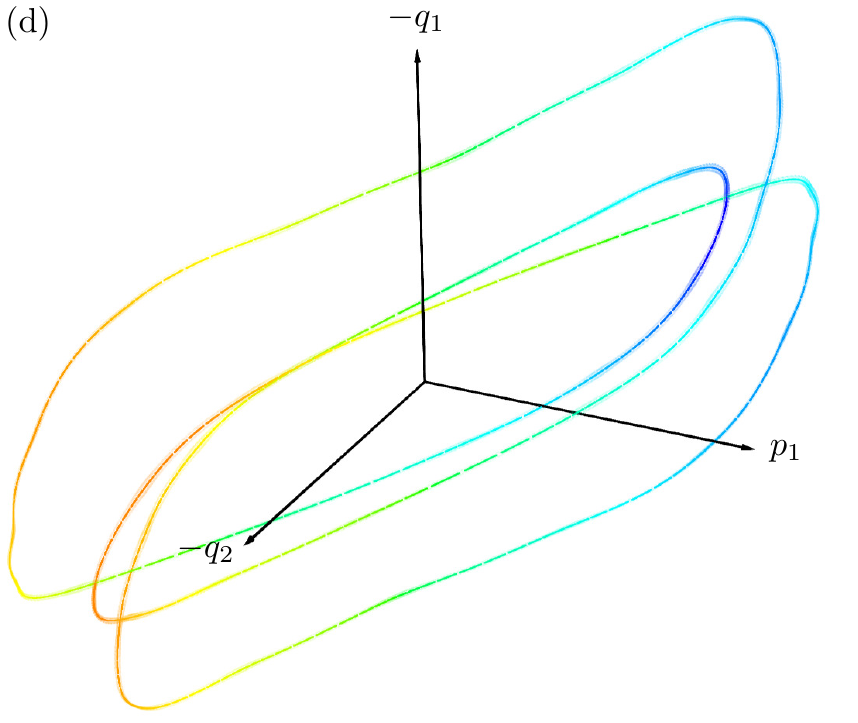}
  \includegraphics{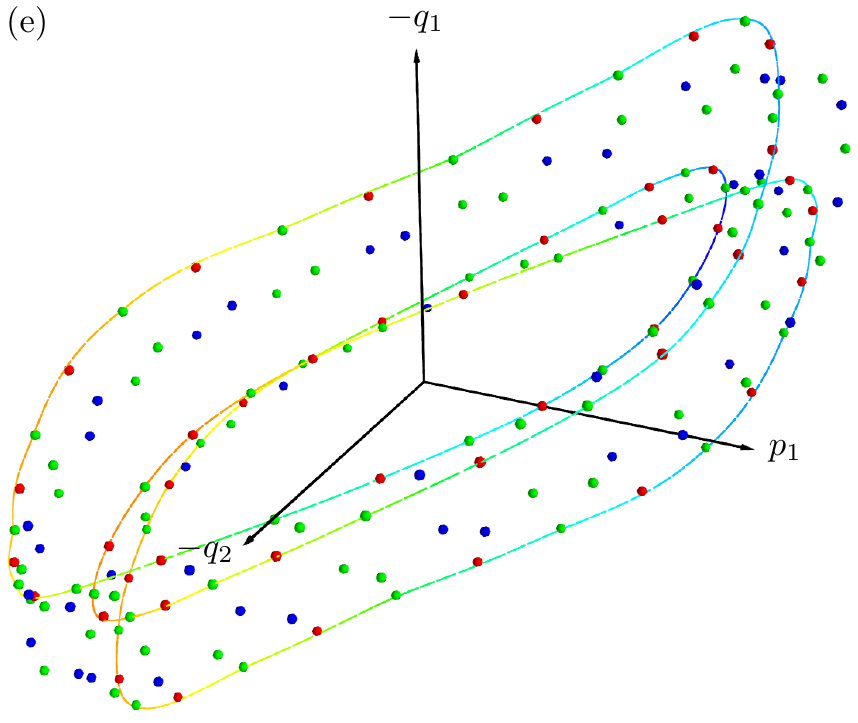}
  \caption{\label{fig:todesco-example} %
    Illustration of the break-up of resonant $2$-tori. \subfiga{}
    Magnification of the frequency plane shown in
    Fig.~\ref{fig:fa-central-tori}. The circled small letters mark the
    frequencies of the example tori for the different cases
    \subfigb{}--\subfige{} and Figs.~\ref{fig:todesco-example2} and
    \ref{fig:todesco-example3}. The example tori are visualized by
    \threeD{} projections as in Fig.~\ref{fig:proj-central-tori}:
    \subfigb{} irrational torus, \subfigc{} elliptic $1$-torus of the
    uncoupled rank-$1$ resonance $7:0:2$ with a $2$-torus from its
    surrounding, \subfigd{} elliptic $1$-torus of the coupled rank-$1$
    resonance $-1:3:0$ with a thin $2$-torus from its surrounding,
    \subfige{} two elliptic-elliptic (red), four elliptic-hyperbolic
    (green) and two hyperbolic-hyperbolic (blue) periodic orbits of
    period $21$ from the rank-$2$ resonance at $(\nu_1, \nu_2) =
    (\nicefrac{2}{7},\nicefrac{2}{21})$. The $1$-torus of \subfigd{}
    is shown for comparison. \MOVIEREF}
\end{figure*}

The break-up of a resonant $2$-torus in a \fourD{} symplectic map
depends on the number $k$ of fulfilled, linearly independent resonance
conditions, Eq.~(\ref{eq:resonance-driven}), denoted as
\emph{rank}~$k$, and whether the resonance is \emph{coupled} or
\emph{uncoupled}. The four possible cases are illustrated by examples
in phase space and the frequency plane, see
Fig.~\ref{fig:todesco-example}. The example tori are chosen as close
as possible to each other such that the geometry of the unbroken tori
resemble each other. Thus, they look like different cases of
resonances for geometrically the same torus.

\paragraph{\label{sec:rank-0}Rank-$0$:} On a non-resonant $2$-torus of
the integrable system each orbit is dense. Such a $2$-torus will in
general survive a perturbation and just be deformed (KAM-theorem). An
example is shown in Fig.~\ref{fig:todesco-example}~\subfigb{} and
marked as point (b) in Fig.~\ref{fig:todesco-example}~\subfiga{}. Such
$2$-tori have been called
\emph{rotational}~\cite{VraBouKol1996,KatPat2011}.

\paragraph{\label{sec:rank-1-uc}Rank-$1$, uncoupled:} If one uncoupled
resonance condition $m_1 \nu_1 = n$ with $m_1 \neq 0$ is fulfilled on
a $2$-torus of the integrable system, each orbit on it densely fills
$m_1$ disjunct lines. At least $2m_1$ of these infinite number of
lines survive a perturbation, alternating between normally elliptic
and normally hyperbolic~\cite{Tod1996}. An example is shown in
Fig.~\ref{fig:todesco-example}~\subfigc{} and marked as point (c) in
Fig.~\ref{fig:todesco-example}~\subfiga{}. The original $2$-torus
broke since it was on the $7:0:2$ resonance leaving a $1$-torus
consisting of seven elliptic lines (hyperbolic lines not shown). This
$1$-torus is displayed in Fig.~\ref{fig:todesco-example}~\subfigc{}
along with a $2$-torus from its elliptic surrounding. In this example
there are another seven elliptic lines (not shown).

\paragraph{\label{sec:rank-1-c}Rank-$1$, coupled:} If one coupled
resonance condition Eq.~(\ref{eq:resonance-driven}) with both $m_1$
and $m_2$ nonzero is fulfilled on a $2$-torus of the integrable
system, each orbit on it densely fills a line. At least one elliptic
and one hyperbolic line survive a perturbation. An example is shown in
Fig.~\ref{fig:todesco-example}~\subfigd{} and marked as point (d) in
Fig.~\ref{fig:todesco-example}~\subfiga{}. The original $2$-torus
broke since it was on the $-1:3:0$ resonance leaving one elliptic line
(hyperbolic line not shown). This $1$-torus and a thin $2$-torus from
its elliptic surrounding are displayed in
Fig.~\ref{fig:todesco-example}~\subfigd{}. Such $2$-tori of rank-$1$
resonances have been called \emph{tube
  tori}~\cite{VraBouKol1996,KatPat2011}.

\paragraph{\label{sec:rank-2}Rank-$2$:} If two independent resonance
conditions Eq.~(\ref{eq:resonance-driven}) are fulfilled on a
$2$-torus of the integrable system, each orbit on it is periodic. If
the resonance is at $(\nu_1=\frac{n_1}{m_1}, \nu_2=\frac{n_2}{m_2})$
then the period is given by $lcm(m_1,m_2)$. At least four periodic
orbits survive a perturbation: either one elliptic-elliptic, one
hyperbolic-hyperbolic, and two elliptic-hyperbolic periodic orbits or
two complex unstable and two elliptic-hyperbolic periodic orbits. Such
a break-up has also been derived from symmetry
considerations~\cite{KooMei1989a} and has been analyzed also for the
case of strong resonances~\cite{GelSimVie2013}. An example for a
rank-$2$ resonance is shown in
Fig.~\ref{fig:todesco-example}~\subfige{} and marked as point (e) in
Fig.~\ref{fig:todesco-example}~\subfiga{}. The original $2$-torus
broke since it was at the intersection of the resonances $7:0:2$ and
$-1:3:0$ leaving two elliptic-elliptic, two hyperbolic-hyperbolic, and
four elliptic-hyperbolic periodic orbits of period $21$. The twofold
number of periodic orbits is analogous to the twofold number of
surviving $1$-tori for the $7:0:2$ resonance. Another example for a
rank-$2$ resonance are the period-$7$ orbits the central lines \Cpo{}
in Fig.~\ref{fig:psslc-central-tori} originate from. These periodic
orbits result from the rank-$2$ resonance at the intersection of the
resonances $7:0:2$ and $3:1:1$, see Fig.~\ref{fig:fa-central-tori}.

\begin{figure}
  \includegraphics{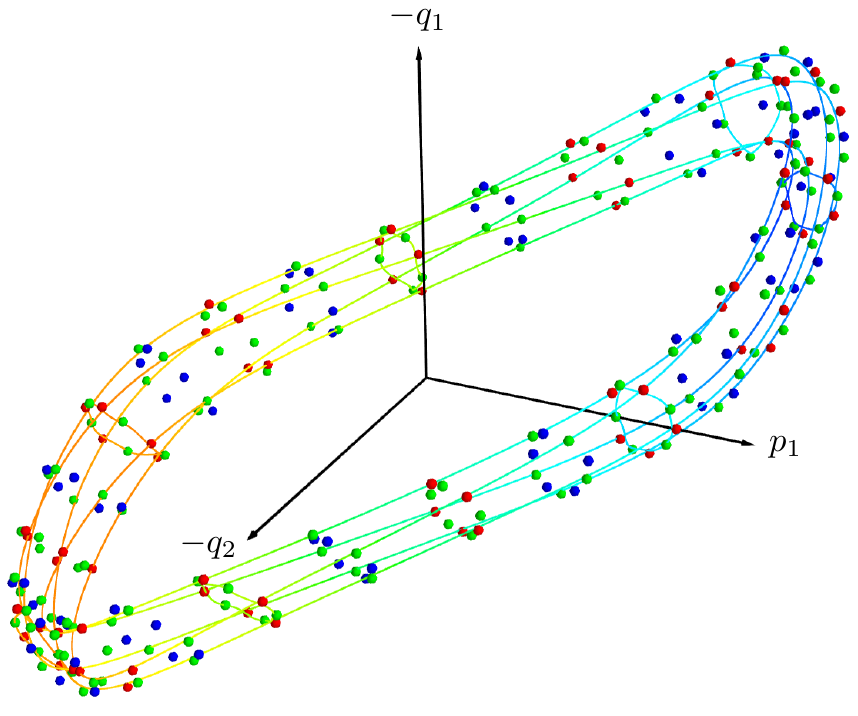}
  \caption{\label{fig:todesco-example2} %
    Rank-$2$ resonance $(\nu_1, \nu_2) =
    (\nicefrac{2}{7},\nicefrac{3}{35})$ at the intersection of the
    resonances $5:-5:1$ and $7:0:2$ marked as point (6) in
    Fig.~\ref{fig:todesco-example}~\subfiga{}. Shown are two
    elliptic-elliptic (red), four elliptic-hyperbolic (green) and two
    hyperbolic-hyperbolic (blue) periodic orbits of period $35$. A
    $1$-torus of the $5:-5:1$ resonance and the $1$-torus of
    Fig.~\ref{fig:todesco-example}~\subfigc{} are shown for
    comparison. \MOVIEREF}
\end{figure}

\begin{figure}
  \includegraphics{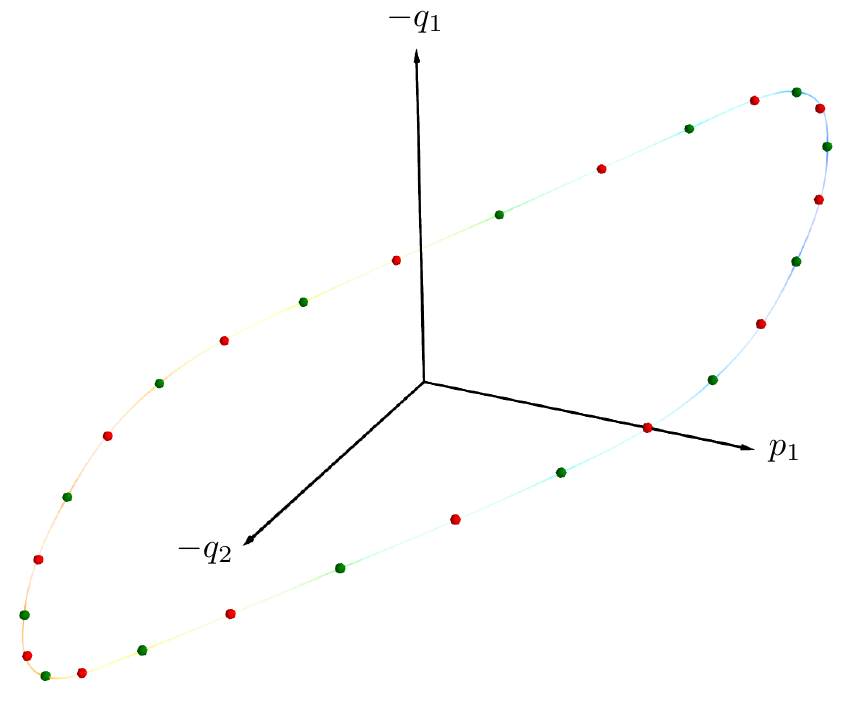}
  \caption{\label{fig:todesco-example3} %
    One-dimensional chain of periodic orbits at the intersection of
    the rank-$1$ resonance $7:0:2$ and the family of elliptic $1$-tori
    \Ma{} marked as point (7) in
    Fig.~\ref{fig:todesco-example}~\subfiga{}. Shown are two
    elliptic-elliptic (red) and two elliptic-hyperbolic (green)
    periodic orbits of period $7$ (also shown in
    Fig.~\ref{fig:proj-central-tori}~\subfiga{}). A $1$-torus of \Ma{}
    close to the intersection is shown for comparison. \MOVIEREF}
\end{figure}

For reasons of continuity we expect the periodic orbits of a rank-$2$
resonance to be located where the surviving lines of the two crossing
rank-$1$ resonances get close to each other in phase space. For
example, we expect an elliptic-elliptic periodic orbit, where an
elliptic line of the first rank-$1$ resonance and an elliptic line of
the second one get close to each other. This is nicely illustrated by
the rank-$2$ resonance in Fig.~\ref{fig:todesco-example2}. There each
of the elliptic $1$-tori of the $5:-5:1$ and $7:0:2$ resonances almost
coincides with a chain of alternating elliptic-elliptic and
elliptic-hyperbolic periodic orbits. Elliptic-elliptic points occur
where these elliptic $1$-tori almost intersect. Note again that there
exists another elliptic $1$-torus from the $7:0:2$ resonance in
between the shown ones.

From this argument also the number and stability of periodic orbits
resulting from a rank-$2$ resonance becomes plausible: The elliptic
($E_1$) and the hyperbolic ($H_1$) line of the first rank-$1$
resonance intersect in four kind of points with the elliptic ($E_2$)
and hyperbolic ($H_2$) line of the second rank-$1$ resonance, giving
periodic orbits with stability $E_1E_2$, $E_1H_2$, $E_2H_1$, and
$H_1H_2$.

As discussed in Secs.~\ref{sec:gaps}~(i) and \ref{sec:hierarchy} a one
dimensional chain of elliptic-elliptic and elliptic-hyperbolic
periodic orbits arises from the break-up of an elliptic $1$-torus with
rational intrinsic frequency. This case is illustrated in
Fig.~\ref{fig:todesco-example3} for the intersection of the $7:0:2$
resonance and \Ma{}.


\end{document}